\documentclass[prl,twocolumn]{revtex4-1}
\usepackage{graphicx}
\usepackage{epsf}
\usepackage{amsmath}
\usepackage{longtable}

\usepackage{amssymb}
\usepackage{color}

\RequirePackage[normalem]{ulem} 
\RequirePackage{color}\definecolor{RED}{rgb}{1,0,0}\definecolor{BLUE}{rgb}{0,0,1} 

\begin{document}

\title{\raggedleft  \sffamily \bfseries \Large Experimental evidence of replica symmetry breaking in random lasers}


\author{ \sffamily \bfseries N. Ghofraniha$^{1,2,3\star}$, I. Viola$^{2,4}$, F. Di Maria$^{5,6}$, G. Barbarella$^5$, G. Gigli$^{4,7}$, L. Leuzzi$^{1,2}$ and C. Conti$^{2,3}$}

\affiliation{  \small \sffamily 
$^1$ Institute for Chemical Physical Processes, CNR (CNR-IPCF), UOS Roma Kerberos, Dept. of Physics, La Sapienza University, Piazzale Aldo Moro 5, I-00185 Rome, Italy.\\
$^2$ Department of Physics, La Sapienza University, Piazzale Aldo Moro 5, I-00185 Rome, Italy.\\
$^3$ Institute for Complex Systems, CNR (CNR-ISC), Via dei Taurini 19, I-00185 Rome, Italy.\\
 $^4$ National Nanotechnology Laboratory  of  CNR-Nanoscience Institute (NNL, CNR-NANO), Via Arnesano, I-73100 Lecce, Italy.\\
$^5$ Institute of Organic Synthesis and Photoreactivity, CNR (CNR-ISOF), Via Piero Gobetti 101, I-40129 Bologna, Italy.\\
$^6$ MIST.E-R Laboratory, Via Piero Gobetti 101, I-40129 Bologna, Italy.\\
$^7$ Department of  Mathematics and Physics "Ennio de Giorgi", Salento University, Via Arnesano, I-73100 Lecce, Italy.\\
$^\star$ Corresponding author: neda.ghofraniha@roma1.infn.it}

\date{\today}

\begin{abstract}
{\sffamily \bfseries
Spin-glass theory is one of the leading paradigms of complex physics
and describes condensed matter, neural networks and biological systems, 
ultracold atoms,
 random photonics, and many other research fields. 
According to this theory, identical systems under
 identical conditions may reach different states.
This effect is known as Replica Symmetry Breaking 
and is revealed by the shape of the probability
 distribution function of an order parameter named the Parisi overlap.
However, a direct experimental evidence in any field of research is still
missing. Here we investigate pulse-to-pulse fluctuations in random lasers, 
we introduce and measure the analogue of the Parisi overlap
in independent experimental realizations of the same disordered sample, 
and we find that the distribution function yields evidence of a transition to a glassy light phase compatible with a replica symmetry breaking.
}\end{abstract}
\maketitle

Replica theory was originally introduced by Edwards and Anderson and detailed by  Sherrington and Kirkpatrick to try to solve the prototype model for spin-glasses~\cite{Mezard}.
It was readily recognized as one of the fundamental paradigms of statistical mechanics and, after Parisi resolution of the mean-field theory~\cite{Mezard}, 
 found  applications in a huge variety of different fields of research~\cite{Amit92,Buc13}. Spin-glass theory gives a rigorous settlement to the physical meaning of  complexity, 
and describes a number of out-of-equilibrium phenomena (for example weak non-ergodicity and aging)
\cite{Mezard,Young}.
More recently this theory has found application in the field of random photonics~\cite{Wie13}, as specifically for random lasers (RLs) and nonlinear waves in disordered systems~\cite{Leu09,Con11,Ant14}.
However, notwithstanding the theoretical relevance, an experimental demonstration of the most important effect, the so-called
replica symmetry breaking (RSB) is still missing. Spin-glass theory predicts that the statistical distribution of an order parameter, the Parisi overlap,
changes shape when a large number of competing equilibrium states emerges in the energetic landscape~\cite{Deb01}.
When this happens, replicas of the system, such as identical copies under the same experimental conditions, may furnish different values of observable 
quantities, because they settle in ergodically separated states after a long dynamics.
 This phenomenon is inherently different from chaos, and relies on
non-trivial equilibrium properties of disordered systems.

RLs are realized in disordered media with gain; the feedback for
stimulated emission of light is given by the scattering and no
external cavity is needed~\cite{Law94}.  Different RLs show multiple
sub-nanometer spectral peaks above a pump threshold~\cite{Cao99}.  The
wide variety of the spectral features reported and the still debated
emission properties are due to the fact that RLs are open systems
where light can propagate in any direction in a disordered fashion
instead of oscillating between well specific boundaries as in standard
lasers~\cite{Vanneste07}.  The scattering strength and pumping
conditions~\cite{Wu08,Leonetti2011,Fol13,Tur10,Bau13} affect the
emission: a large number of modes may result in a continuous
broadband, or exhibit distinguishable resonances. \\ Various authors
have reported evidences of shot-to-shot fluctuations in RLs.
In the study by Fallert {\it et al.}~\cite{Fal09} they are ascribed to the coexistence of localized and
extended modes without analyzing the statistical distribution of their
shot-to-shot correlations.  In~\cite{Van06} they are due to the
intrinsic matter fluctuations in fluids, that are not present in
samples with static disorder as in the case here considered.
In the study by Lepri {\it et al.}~\cite{Lep07} the analysis of the different diffusion regimes is
reported and it is not linked to a phase transition from spontaneous
emission to RL, because the exponent of the diffusion law does not
help to discriminate lasing from fluorescence.  In the study by Mujumdar {\it et al.}~\cite{Muj07}
mode-competition and seemingly chaotic behavior have been considered,
and intended as a generic sensitivity to initial conditions, no
analysis of the distribution of the shot-to-shot correlations is
provided.  In the study by Skipetrov {\it et al.}~\cite{Ski2000} fluctuations due to a transparent Kerr
medium are theoretically considered, and even if they are not strictly
related to the case with gain and loss of RLs, they look to have
connections with the spin-glass approach in the study by Leuzzi {\it et al.}~\cite{Leu09,Con11,Ant14}.
The latter approach has been developed in a series of papers, and is
based on the mode-coupling laser theory as, for example,
in the study by Leonetti {\it et al.}~\cite{Leo13}, which reports on the time dynamics of frequency
locking mechanism and without addressing the statistical properties of
the laser emission.  Our understanding is that the RL spectral
fluctuations are a manifestation of the existence of many degenerate
lasing states, each one corresponding to a given ensemble of activated
modes with their own wavelengths, phases and intensities.  This theory
is a statistical mechanical formulation obtained starting from the
decomposition in normal modes of the electromagnetic field of light,
in the presence of non-linearity, and leading to a quenched disordered
spin-glass-like Hamiltonian.  The theoretical approach considers the
combined effect of localized and radiating modes as occurring in any
open cavity, and allows to describe open systems with arbitrary degree
of openness \cite{Ant14}.  A complex landscape typical of glassy
systems is expected to occur beyond some threshold critical value of
the external parameter driving the transition. The measurements
reported in this manuscript confirm this theory: in a system that 
might be properly represented by a spin-glass theory -
Refs.~\cite{Leu09,Con11,Ant14} - we measure the spin-glass order
parameter distribution function, such as $P(q)$ and we find a behavior
akin to the one theoretically representing the spin-glass phase (at
high pumping) and the paramagnetic/fluorescence phase (at low
pumping), as well as the transition between them.

\section{\bf Results}
\noindent {\bf Spin-glass theory of random lasers}

 \noindent In the following, we report on the shot-to-shot emission
 fluctuations from planar RLs made of a fluorescent $\pi$-conjugated
 oligomer in amorphous solid phase and we analyze the experimental
 results by means of the replica theory.

At variance with standard chaotic ordered lasers~\cite{Gio83,Hal83},
in which, for specific and tailored conditions, few modes provide
exponentially diverging temporal trajectories, RLs are thermodynamic
systems, with thousands of modes and degrees of freedom, which
exhibits a huge number of (meta-)stable states.  A standard chaotic
laser will always display the same spectral behaviors in the same
conditions; at large variance with what we observe in RLs.

In the spin-glass approach the RL modes are treated  as continuous complex spin variables with a global (power) constraint and whose coupling is governed by the interplay between disorder and nonlinearity.
We remark that in this theoretical analysis, the Hamiltonian description is an effective one, 
representing the stationary regimes under pumping. The role of inverse temperature in equilibrium 
thermodynamics is played by the energy pumped into the system. The Hamiltonian description is derived in such a way to encode, 
in a distinct, effective temperature space, pumping and dissipation of the laser. Hence, the theoretical approach 
also includes the fact that RLs are open dissipative systems with an external pumping mechanism and radiation losses~\cite{Ant14}.
In our experiments the disorder is fixed and the nonlinearity increases with the pumping energy, 
which acts as the inverse of temperature in statistical mechanics~\cite{Wei10,Gordon02}: 
at low energy  (high temperature) there is no gain competition~\cite{Ram12,Tur08} between the modes
  and they oscillate independently 
in a continuous wave paramagnetic regime;
while at high energy (low temperature) 
the coexistence of mode coupling through the gain competition and  frustration due to  disorder gives rise to a glassy regime. 
In this regime a large number of electromagnetic modes is activated and in interaction.
 The set of the activated mode configurations 
is found to change from pulse-to-pulse. 
We will refer to each set of configurations as a state.
This is justified by the fact that during 
a single pulse of the pump beam, very many stimulated
emission processes take place at each mode frequency, that is, 
each mode performs a long dynamics, compatible with thermalization. 

The observation 
of numerous different states can be understood as the evidence of a 
 thermodynamic phase  corresponding to
many valleys separated by barriers in the corrugated free energy landscape.
The same sample under identical experimental conditions  furnishes different laser emissions. Each different instance
 of laser emission is, thus, a physical realization of a replica, in the meaning introduced in replica theory~\cite{Mezard}.\\

\noindent {\bf Random laser emission}

\begin{figure*}[h!]
\includegraphics[width=0.8\textwidth]{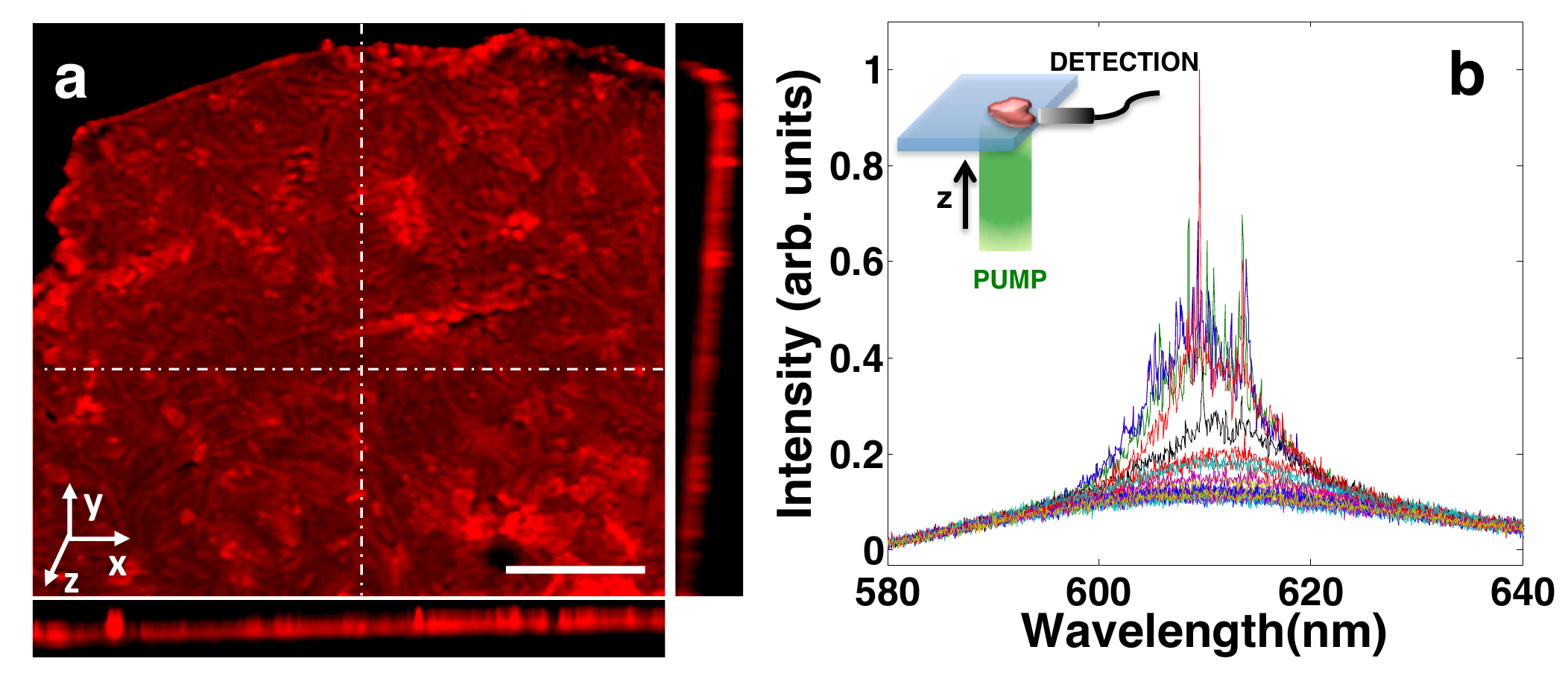}
 \caption{  
{\bfseries\sffamily Sample image and emission spectra showing random lasing.} {\sffamily(a) 3D-reconstruction of confocal microscopy $z$-stack images of the supra-molecular laminar packing in aT5OCx solid sample. 
The right and the bottom panels report the $yz$- and the $xz$- sections, respectively. Scale bar: 20$\mu$m.
(b) High resolution single shot spectra taken in the same conditions, 10mJ pump energy. Inset: sketch of the experiment.} }
\label{fig1}
\end{figure*}

\noindent The investigated system is a functionalized thiophene based oligomer commonly named T5OCx (see Methods for details) in amorphous solid state, as shown in the
confocal microscopy image reported in  fig.~\ref{fig1}a .
 RL in T5COx  spin-coated in thin films~\cite{Anni04} and lithographed in micro-structures~\cite{Gho13}  has been previously observed.
The strong   density  of the T5OCx supra-molecular laminar packing in the solid samples allows to study shot-to-shot emission fluctuations not  reported so far.    
  A sketch of the pumping and collecting geometry is given in the inset of fig.~\ref{fig1}b. Experimental  details are reported in Methods. 

\begin{figure*}[h!]
\includegraphics[width=0.8\textwidth]{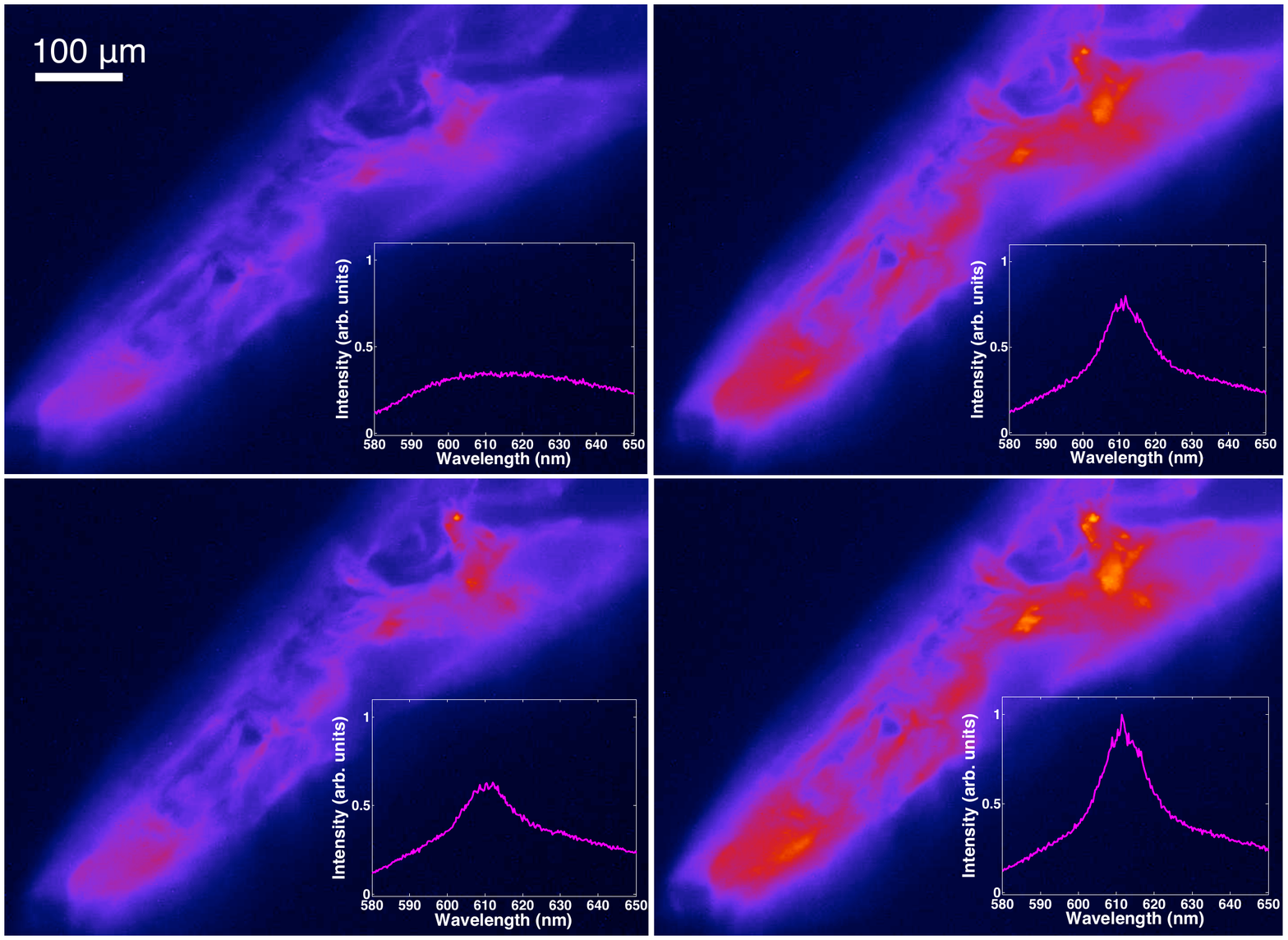}
 \caption{ 
{\bfseries\sffamily Snapshots of RL emissions.}
{\sffamily Single shot optical images and corresponding emission spectra (insets) during the pumping of the sample in the same experimental conditions. The input energy is 10mJ. } }
\label{fig2}
\end{figure*}

 We illustrate in fig.~\ref{fig1}b single shot high resolution (0.07nm) emission spectra,
taken at identical experimental conditions. Input energy is 10mJ. 
The presence of  RL modes with configuration variable  from pulse-to-pulse is evident:
each time the system is pumped the numerous passive modes randomly compete  for  the available gain,
giving rise to several different compositions of the activated spectral peaks~\cite{Muj07}. 
Such behavior is also evidenced by the direct visualization of the sample during pumping, as  reported in fig.~\ref{fig2}, where four different 
fluorescence images 
taken at four single shots  exhibit different emission patterns and corresponding spectral features.
It is clear that from
shot to shot the intensity and the profile of the bright spots, where
the  RL modes are concentrated, change. Each bright spot contains many
interacting modes as evidenced by the spectra with many peaks and by the size of
the spots.  Many extended
modes are concentrated in different extended spots and from shot to shot
the size and the brightness of these spots change showing that the
spatial structure of the modes changes.

\begin{figure*}[h!]
\includegraphics[width=0.8\textwidth]{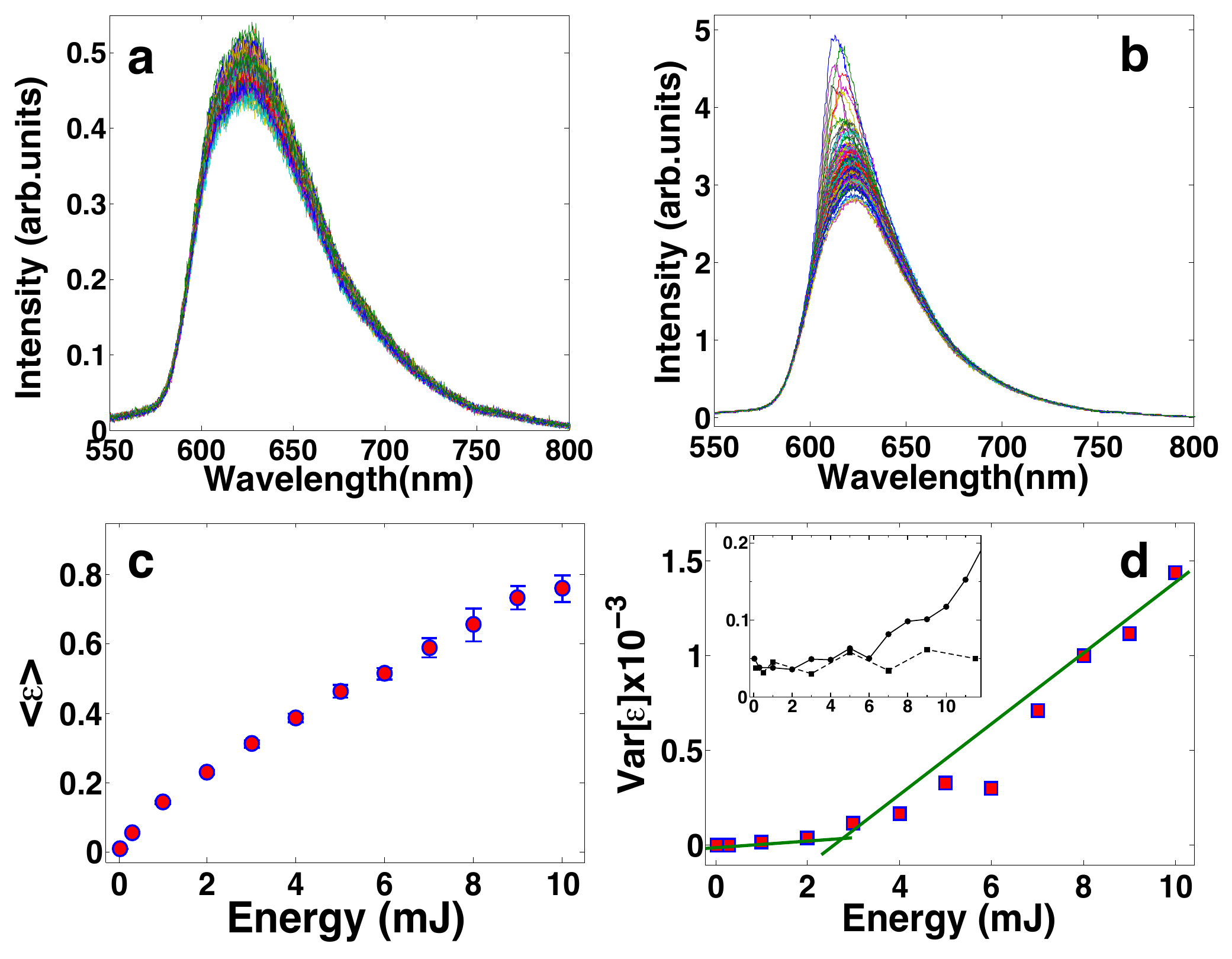}
 \caption{ 
 {\bfseries\sffamily  Fluctuations  in the random laser emission. } 
{\sffamily a-b) Emission spectra at low energy 1mJ (a) and high energy 10mJ (b). 
c) Mean emission intensity $<\epsilon>$ vs. pump energy, error bars are  standard deviation. d) Variance of $\epsilon$ vs. pump energy.
Inset: peak standard deviation divided by mean peak intensity from the sample (solid line) and the pump laser (dashed line) emission spectra.} }
\label{fig3}
\end{figure*}

 In figs.~\ref{fig3}a and~\ref{fig3}b we show emissions of subsequent 100 shots with lower spectral resolution (0.3nm), at two different pump energies.
There is no time periodicity in the spectral fluctuations: their sequence  is random. 
At low energy the noisy variations of the spontaneous emission are negligible if compared to the pronounced fluctuations  observed at high energy.\\

\noindent {\bf Analysis of the spectral emissions}

\noindent 
The cavity coupled modes in the Hamiltonian describing random lasers~\cite{Leu09} are variables not easily accessible in
the experiments.Ê We can otherwise measure spectra and have
access to the intensities  $I(k)\propto \langle |a_{j}^2| \rangle_k$, 
where
$a_{j}$ is the amplitude of the longitudinal mode $j$~\cite{Leu09,Con11} 
at wavelength $\lambda_j$.
 In our analysis each spectrum represents a different state
of the same thermodynamic phase. The average is taken over all modes
having similar wavelength, binned together in data acquisition with index $k$, due to the finite spectral resolution $\Delta \lambda = 0.3$ nm.
We define the average spectral intensity as
\begin{equation}\label{epsilonalpha}
\epsilon_\alpha=\frac{1}{N}\sum_{k=1}^{N} I_{\alpha}(k),
\end{equation}
being $\alpha$  the
 replica index and $N$ the number of acquired spectral points.   
The variation of this parameter from shot-to-shot quantifies the observed fluctuations.  
Mean and variance of $ \epsilon_{\alpha}$ over $N_s$ replicas for each pumping are
 \begin{equation} \label{epsilonmean}
<\epsilon>=\frac{1}{N_s}\sum_{\alpha=1}^{N_s} \epsilon_{\alpha},
\end{equation}
\begin{equation} \label{epsilonvar}
\rm{Var}[\epsilon]=\frac{1}{N_s}\sum_{\alpha=1}^{N_s} (\epsilon_{\alpha}-<\epsilon_{\alpha}>)^2.
\end{equation}

Figure~\ref{fig3}c shows the continuous growth
 for increasing input energy of the mean intensity,  Eq.(\ref{epsilonmean}).
 It is very important to explain the stimulated emission we are observing.
 In our samples
we see the coexistence of RL and   amplified spontaneous emission (ASE)~\cite{Anni04, Fro99}.
The latter is evident from the absence of threshold in the spectral peak intensities in fig.~\ref{fig3}c and RL is evident from the fine peaks 
superimposed to the ASE band in fig.~\ref{fig1}b.
This specific system,  made of thick  amorphous solid dye, displays a threshold in terms of the spectral fluctuations,
quantified by $\rm{Var}[\epsilon]$, calculated as  Eq.~(\ref{epsilonvar}) and shown in fig.~\ref{fig3}d. Such behavior has not been reported in  previous works on similar dyes \cite{Anni04, Fro99}.
 It is well-known that ASE does not exhibit a threshold, at variance with RL. However, in our case the RL threshold 
 is masked by ASE when coming at the spectral peak. 
 On the contrary, the threshold of the fluctuations in fig.~\ref{fig3}d
 reveals the onset of the RL action, that for this sample is estimated around 3mJ.
 All experiments are performed in the same environmental conditions, that is the temperature, the position and the state of the sample 
 and the performances of the optical components  are controlled to be stable by growing pump energy. 
 To check the stability of the input laser, we estimate the energy fluctuations as the percentage variation given by  the ratio between 
 the standard deviation and the mean value of the emission peak intensity with and without sample, for increasing energy. 
 The plots are reported in the inset of fig.~\ref{fig3}d, where the laser energy variations 
 are approximately around 4\% at any energy, 
 while the same quantity from the sample  increases at high energy. We stress that an error of 4\% on the x-axis of fig.~\ref{fig3}d is 
 inside the symbols and negligible compared to the energy values, thus it cannot cause the evident increase of the observed variance.
 
To analyze the system behavior in the framework of  statistical mechanics of disordered systems, and
characterize the high pumping RL regime, we introduce another order parameter to identify the phase transition:
the overlap between intensity fluctuations in different experimental replicas. 
The experimentally accessible variable, coarse graining the behavior of single modes, is the intensity fluctuation at a given frequency:
\begin{equation}
 \Delta_\alpha(k)=I_{\alpha}(k)-\bar I(k)
\end{equation}
with $\bar I(k)$ the average over replicas of each  mode intensity
\begin{equation}
\bar I(k)=\frac{1}{N_s}\sum_{\alpha=1}^{N_s}I_{\alpha}(k).
\end{equation} 
The overlap between pulse-to-pulse intensity fluctuations is defined as
\begin{equation}\label{overlap}
q_{\alpha\beta}= \frac{\sum_{k=1}^N \Delta_\alpha(k)  \Delta_\beta(k)}{\sqrt{\sum_{k=1}^N \Delta^2_\alpha(k) } \sqrt{\sum_{k=1}^N \Delta^2_\beta(k) }}.
\end{equation} 

From the measured spectra we calculate the set of all $N_s(N_s-1)/2$ values $q$ of $q_{\alpha\beta}$  for  each different input  energy,
determining  their distribution $P(q)$. 
\begin{figure*}[h!]
\includegraphics[width=0.6\textwidth]{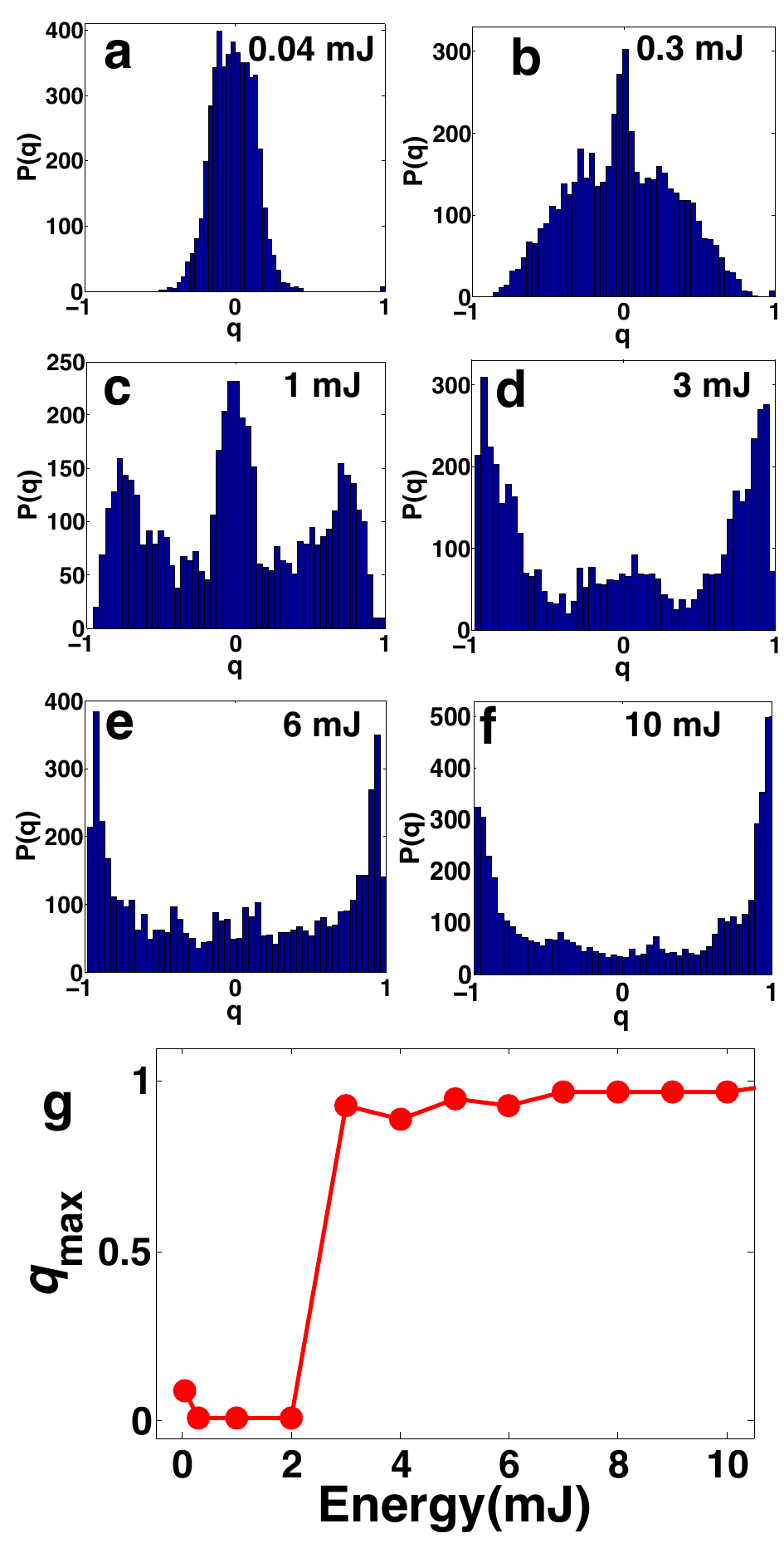}
 \caption{
 {\bfseries\sffamily Distribution function of the overlap showing replica symmetry breaking by increasing pump energy.}
{\sffamily a-f) Distribution of the overlap $q$ at different pump energy. g) $q_{\rm max}$ corresponding to the position of the maximum of $P(|q|)$
 versus pumping. }
} 
\label{fig4}
\end{figure*}
We retain $P(q)$ as a measure of the theoretical distribution of the overlap between 
mode amplitudes \cite{Leu09,Con11},
which is a leading quantity for the description of glassy phases~\cite{Mezard}.
So far no experimental measurements of $P(q)$ have been reported.   

Six examples are reported in fig.~\ref{fig4}a-f for increasing pump
energy. At low energy (fig.~\ref{fig4}a), all overlaps are centered
around the zero value, meaning that the electromagnetic modes are
independent and not interacting in the paramagnetic regime.  By
increasing energy, modes are coupled by the nonlinearity and this
corresponds to a non-trivial overlap distribution
(fig.~\ref{fig4}b-f).  In the high energy glassy phase, with all modes
highly interacting and frustrated by the disorder, $q$ assumes all
possible values in the range [-1,1].  Such behavior of the $P(q)$
evidences the fact that the correlation between intensity fluctuations
in any two replicas depends on the replicas selected. The variety of
possible correlations extends to the whole range of values. This is a
manifestation of the breaking of the replica symmetry. In fig.~\ref{fig4}g we
show $q_{\rm max}$ corresponding to the position of the maximum of
$P(|q|)$ versus pumping: it changes drastically signaling a phase
transition between 2mJ and 3mJ, compatible with the threshold
determined by Var[$\epsilon$] in fig.~\ref{fig3}d.  We point up that in
the high pump regime $P(q)>0$ for any value of $q$.

\begin{figure*}[h!]
\includegraphics[width=0.8\textwidth]{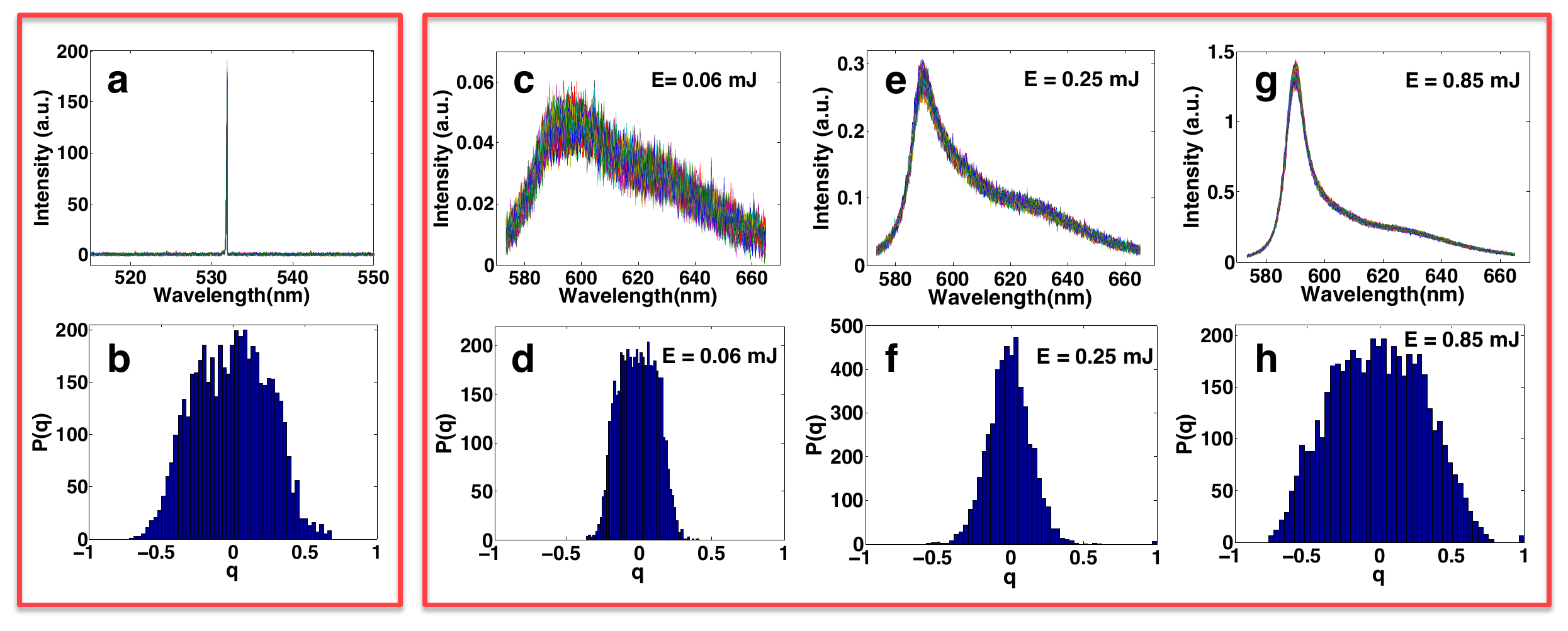}
 \caption{ 
 {\bfseries\sffamily   Distribution function of the overlap in standard and random laser without replica symmetry breaking. } 
{\sffamily a-b) Emission spectra (a) and  $P(q)$ (b) of a Q-switched pulsed Nd-Yag standard ordered laser.  The analysis is done on 100 shots.
c-h)  Emission spectra (c,e,g) and correspondent $P(q)$ (d,f,h) of a liquid dispersion of titanium dioxide in Rhodamine B-ethylene glycol solution at three different pump energy through the threshold.} }
\label{fig5}
\end{figure*}

To stress the peculiarity of such a signature of RSB and the relevance of the analysis method proposed in probing the onset of a prominent 
glassy nature of RL regimes in different compounds, we compare the different behavior of $P(q)$'s
generated by an identical analysis for the emission spectra of a standard ordered laser, the same used as pump laser in the experiments,
as reported in fig.~\ref{fig5}a, and of a differently structured RL, a liquid dispersion of titanium dioxide in 3mM Rhodamine B-ethylene glycol solution,  shown in fig.~\ref{fig5}c,e,g. 
In fig.~\ref{fig5}b the standard laser (obviously well above
threshold) does not show any symmetry breaking, even though small 
fluctuations are present, as expected in an ordered system.
In fig.~\ref{fig5}d,f,h the $P(q)$'s for the 
liquid RL are displayed  by increasing pump energy across its estimated threshold
of 0.25mJ and no evident signs of replica symmetry breaking above the lasing threshold can be appreciated up to 0.85 mJ.
This result points out that indications about the occurrence of RSB depend on the characteristics of the system and are, therefore, a way to classify
the nature of different RLs.

\section{\bf Discussion}
 We experimentally show that the RL intensity fluctuations in a solid disordered system grow
 drastically as the pump energy increases beyond a threshold,
 exhibiting a transition from negligible to large oscillations.  A
 possible order parameter is quantified by the variance of the
 fluctuations.  A more refined one is the distribution of all mutual
 overlaps between intensity fluctuations in different physical
 replicas. Besides evidencing the threshold of phase transition, it
 describes the organization of states and their non-trivial
 correlation, non-symmetric in replicas exchange.  Above the threshold
 pump energy, the overlap between two replicas does depend on the
 specific couple chosen and the variation from couple to couple turns
 out to be more complicated than the ordinary statistical
 fluctuations. That is, replicas are not all equivalent to each other
in the analyzed random laser
 and we, thus, observe an instance of replica symmetry breaking
 through the random laser transition.\\

The parameter $q$ measures the overlap of the shot-to-shot fluctuations at each wavelength around the average spectrum at a given pumping. 
Thus $q\sim 1$ when two shots both have very similar intensity fluctuations at each wavelength. 
In principle, it does not matter how large the fluctuations are to have a large $q$, but only how much they are correlated. 
The fact that large overlap values become probable in the lasing regime above threshold is not a trivial feature. 
Similarly, one has $q\sim -1$ when two shots display spectral fluctuations similar in magnitude (no matter how large) but opposite in sign, that is the 
shot-to-shot fluctuations around the average spectrum are very anti-correlated. When $q\sim 0$ the intensity fluctuations in two shots are basically uncorrelated.
This can imply the ASE background, but can also correspond to the presence of a sub-set of shots with uncorrelated large fluctuations. 
Above threshold, the latter situation occurs, as can be observed looking at the single spectra. 
By the definition of overlap in Eq.~(\ref{overlap}) in all cases $|q|\sim[0,1]$, the fluctuations of a single spectrum can be large or small. 
The important observation is that increasing the pumping, the correlations between shot-to-shot fluctuations
 become non-trivial right when the fluctuation magnitude starts increasing significantly (such as above threshold), see fig.~\ref{fig3}d.

In the present work not only a first experimental proof of RSB is presented, but
a new method is provided that allows to characterize the nature of the randomness above the lasing threshold applicable to the whole profusion of random lasers.
By directly probing the measure of spectral intensity fluctuations, 
the possible existence and the organization of many equivalent competing lasing states  yielding a glassy light behavior can be verified, allowing for an alternative, supplementary RL classification in terms of glassiness. 

As a last remark, we recall that in the effective statistical
 mechanical Hamiltonian for
RLs the variables are, actually, the modes, such as complex
 amplitudes~\cite{Leu09,Con11,Ant14}, not easily accessible in
  the experiments. Whereas we are considering their squared
 modulus given by the emitted intensity, thus losing information about the actual state structure.
 Looking at the present results we stress, however, that such coarse-graining appears to be refined
 enough to investigate the peculiar  properties of glassy-like regimes of random lasers.

{\section*{ \large\sffamily \bfseries Methods}
{\footnotesize  
\noindent{\bfseries Sample preparation and characterization.} 
The investigated disordered system is a functionalized thienyl-S,S-dioxide quinquethiophene (T5OCx), 
whose preparation 
modalities and molecular structure are reported elsewhere~\cite{Gho13}, in amorphous solid state obtained by the crystallization of the dye powder after 
a thermal shock above the  transition temperature ($T_{\rm g}$). Successively, the packing states have been frozen by cooling across the $T_{\rm g}$.
The thermal process confers to the $\pi$-conjugated oligomer a crystal-like molecular ordering over large area with 
the appearance of laminar domains that grow anisotropically in time up to their final random packing
 leading, at the end of the process, to a macroscopic disordered system  as shown in fig.~\ref{fig1}a.
 Sample characterization is performed by a $z$-stack sequential acquisition of confocal microscope (Olympus, FV1000). 
 The optical sections are collected in transverse $xz$- and $yz$- planes.
 The $x$-$y$ surface  area of  the investigated samples is in the range ($10^5$-$10^6)\mu m^2$ and the average thickness along $z$ is (7$\pm$2)$\mu$m. \\

\noindent{\bfseries Measurements.} 
The samples are pumped by a frequency doubled Q-switched  Nd:YAG pulsed laser emitting at $\lambda$=532~nm, 
with 10~Hz repetition rate, 6~ns pulse duration and with 12~mm beam diameter, the emitted radiation is 
collected from the sample edge  into an optic fiber connected 
 to a  spectrograph equipped with electrically cooled CCD array detector, with
gratings density of 600 mm$^{-1}$ and 1800 mm$^{-1}$, for low (0.07nm)
and high (0.3nm) resolution spectra respectively. 
 The samples are put on glass slides and pumped perpendicularly along the $z$ direction and 
 all reported spectra are taken from single pumping shots.

 }

\section*{\large\sffamily \bfseries References}

\section*{\large\sffamily \bfseries Acknowledgements}

The research leading to these results has received funding from
 the Italian Ministry of Education, University
and Research under the Basic Research Investigation
Fund (FIRB/2008) program/CINECA grant code
RBFR08M3P4 and under the PRIN2010 program, grant code 2010HXAW77-008 and 
from  the People Programme (Marie Curie Actions) of the European Union's Seventh Framework Programme FP7/2007-2013/ under REA grant agreement n¡ 290038, NETADIS project.\\

\section*{\large\sffamily \bfseries Statement of authors contribution}
N. G. and C. C. worked on the experimental characterization of the random laser emission.
I. V., F. D.M., G. B. and G. G. worked on the sample preparation and characterization.
L. L. worked on the theoretical interpretation of the data.
N. G., L.L., and C.C. wrote the manuscript.

  \end{document}